\documentclass[%
reprint,
superscriptaddress,
amsmath,amssymb,
aps,
prb
floatfix,
showpacs,
]{revtex4-2}

\usepackage{graphicx}
\usepackage{dcolumn}
\usepackage{bm}
\usepackage{hyperref}
\usepackage{ulem}

\usepackage[english]{babel}
\usepackage[utf8]{inputenc}
\usepackage{amssymb,amsfonts,amsmath,mathtools,mathrsfs}
\usepackage{relsize}
\usepackage{mdframed}
\usepackage[margin=0.6in]{geometry}
\usepackage{enumitem}
\usepackage{graphicx}
\usepackage{float}
\usepackage{url,hyperref}
\usepackage[table,usenames,dvipsnames]{xcolor}
\usepackage{gensymb}
\usepackage{grffile}

\hypersetup{
	colorlinks=true,
	linkcolor=Blue,
	filecolor=magenta,      
	urlcolor=Blue,
	citecolor=Blue,
}
\usepackage[bottom]{footmisc}
\usepackage[capitalize]{cleveref}
\usepackage{textcomp}
\usepackage{braket} 

\setcitestyle{line}

\usepackage{amsmath,amssymb,wasysym}

\makeatletter
\def\@fnsymbol#1{\ensuremath{\ifcase#1\or *\or \dagger\or \ddagger\or
   \mathsection\or \mathparagraph\or \|\or **\or \dagger\dagger
   \or \ddagger\ddagger \else\@ctrerr\fi}}
    \makeatother

\begin{document}

\title{Competing interaction driven broad magnetic entropy distribution in triangular net compounds \text{Ba$_3$RE(BO$_3$)$_3$},  RE= Dy and Ho}

\author{Krishna Nand Prajapati}\email[E-mail:] {knpiitb@gmail.com}
\address{Department  of Physical Sciences, Indian Institute of Science Education and Research, Berhampur, Odisha 760003, India}

\author{Prabuddha Kant Mishra}
\address{Central Advanced Instrumentation Facility, Indian Institute of Science Education and Research, Berhampur, Odisha 760003, India}

\author{Shashi B Singh}\email[E-mail:] {sbsingh@iiserbpr.ac.in}
\address{Department  of Physical Sciences, Indian Institute of Science Education and Research, Berhampur, Odisha 760003, India}

\begin{abstract}

Triangular magnetic lattices have attracted significant attention as a model platform for exploring frustrated magnetism and quantum spin liquids, particularly in rare-earth-based systems. These materials also hold considerable promise for cryogenic magnetic refrigeration. In this work, we investigate the magnetic and magnetocaloric properties of the rare-earth (RE) based compounds $\text{Ba}_3\text{RE}(\text{BO}_3)_3$ ($\text{RE} = \text{Dy}$ and $\text{Ho}$). Polycrystalline samples synthesized via a solid-state reaction method crystallize in a hexagonal structure with the space group $P6_3mc$, featuring triangular nets separated by approximately $8.8\text{ \AA}$ along the $c$-axis. Detailed magnetic investigation reveals a deviation from pure paramagnetic behavior at low temperatures ($T < 20\text{ K}$). This is further corroborated by isothermal magnetization measurements, which exhibit non-overlapping $M$ vs. $H/T$ plots and reduced magnetization moments, suggesting significant crystal electric field (CEF) splitting of localized $f$-states. For an applied magnetic field change of $70\text{ kOe}$, the maximum entropy change (Relative Cooling Power) reaches $4.2\text{ J kg}^{-1}\text{ K}^{-1}$ ($204\text{ J kg}^{-1}$) and $7.3\text{ J kg}^{-1}\text{ K}^{-1}$ ($230\text{ J kg}^{-1}$) for the Dy- and Ho-based compounds, respectively. Overall, the positive slopes of Arrott plots, the universal scaling behaviour of $\Delta S_{\text{m}}/\Delta S_{\text{m}}^{\text{max}}$ versus reduced temperature ($t$), and the field exponent $c$ of the RCP versus $H$ curves indicate a second-order nature of entropy distribution, making them promising candidates for magnetocalorics in the millikelvin range.
\end{abstract}


\maketitle


\section{Introduction}

Conventional ultra-cooling technology is predominantly based on the vapour compression cycle, which is a major contributor to global warming and related environmental issues \cite{phan2007, franco2018magnetocaloric}. Many alternative refrigeration techniques have been explored to overcome this limitation, based on electrocaloric, mechanocaloric, and magnetocaloric refrigeration \cite{kitanovski2020,manosa2017_machanocloric,griffith2018_magnetocaloric}. Recently, the magnetocaloric effect (MCE) has attracted huge attention as a promising route toward solid-state refrigeration. The  MCE is a magneto-thermodynamic phenomenon that induces a change in the magnetic entropy of a material by application or removal of an external magnetic field, resulting in the corresponding temperature variation under adiabatic conditions \cite{griffith2018_magnetocaloric}. The nature of magnetic phase transitions (MPTs), as well as the geometric frustration of the material, strongly influence the MCE performance. The first-order MPT material produces a larger MCE but often suffers from thermal and hysteresis losses, reducing the refrigeration efficiency compared with the second-order MPT materials \cite{Gutfleisch_hystersis}. Therefore, second-order MPT materials with larger MCE are a robust candidate for magnetic cooling where the system has to pass through repeated cycles \cite{li2012_MPT}. Further, geometrical frustration of materials can suppress conventional long-range ordering and promote complex spin configurations, leading to interesting low-temperature magnetocaloric responses \cite{Mishra2024a, Mishra2024}. 

Identifying suitable materials with high magnetocaloric performance is crucial for practical magnetocaloric applications. Several materials, including intermetallic alloys, rare-earth (RE)-based perovskites, spinel ferrites, and RE-containing magnetic alloys or compounds, have been explored as magnetocaloric materials \cite{Mishra2024, Mishra2024a, Rooney2026MgLnB5O10,islam2020}. Recently, magnetically frustrated triangular-net REs in various oxides have attracted significant attention for higher MCE performance. These are promising materials for MCE due to ease of synthesis, low-cost base materials, and relatively greater chemical stability \cite{Guo2019KBaREBO3,Kelly2025Tb,klinger2025_REBaBO3,ennis2024_Ba3ErBO3,Gao2018}.   

\begin{figure*}[t]
\includegraphics[width=0.8\linewidth]{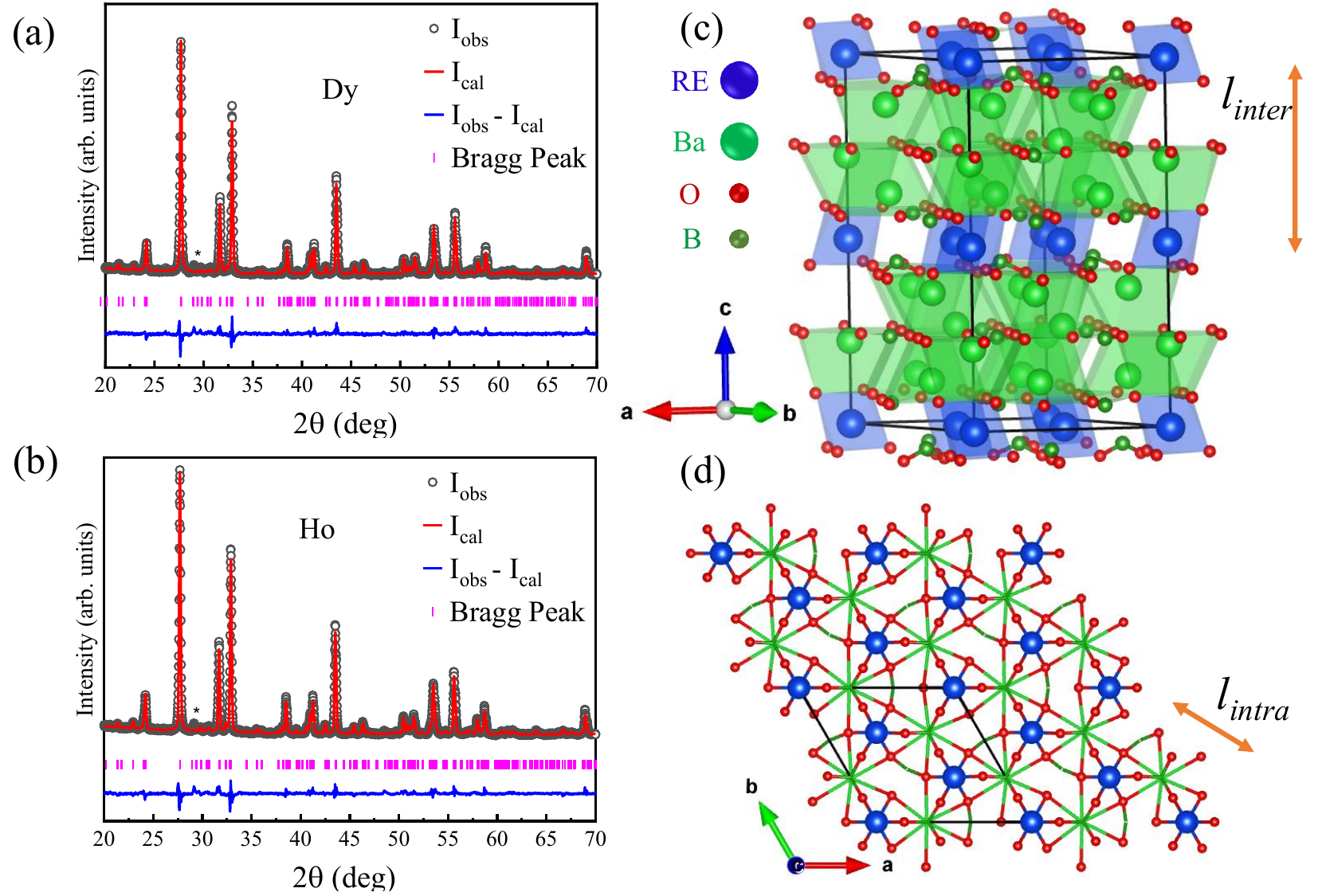}
\caption{(a), (b)} Rietveld refinement of the X-ray diffraction (XRD) pattern for \textbf{(a)} Ba$_3$Dy(BO$_3$)$_3$ and \textbf{(b)} Ba$_3$Ho(BO$_3$)$_3$. The black open circles represent the observed intensity ($I_{\text{obs}}$), the solid red line denotes the calculated profile ($I_{\text{cal}}$), and the solid blue line at the bottom represents the difference curve ($I_{\text{obs}} - I_{\text{cal}}$). The vertical magenta tick marks indicate the Bragg peak positions. \textbf{(c), (d)} Schematic representations of the crystal structure showing atomic packing, coordination polyhedra viewed along different crystallographic projections, and the spatial parameters $l_{intra}$ and $l_{inter}$, which represent intraplane and interplane nearest-neighbour metal-ion distances.
\label{Str}
\end{figure*}

In this work, we present a detailed investigation of the magnetic and magnetocaloric properties of the rare-earth (RE) borates $\mathrm{Ba_{3}RE(BO_{3})_{3}}$ ($\mathrm{RE = Dy, Ho}$). The crystal structures of these materials are well known \cite{Cox1994Ba3MBO3}, but detailed understanding of low-temperature magnetism on the frustrated triangular lattice remains largely unexplored. A recent study on a related member of this family ($\mathrm{RE = Er}$) has revealed competing two-sublattice exchange interactions, giving rise to complex coexisting ferromagnetic and antiferromagnetic correlations \cite{ennis2024_Ba3ErBO3}. Furthermore, the $S = 1/2$ ytterbium analogue $\mathrm{Ba_{3}Yb(BO_{3})_{3}}$ has been examined via nuclear magnetic resonance (NMR) spectroscopy down to $T = 0.26\text{K}$, where it displays strong quantum spin fluctuations without any indication of long-range ordering or spin freezing \cite{bag2021_Ba3YbBO3, Zeng2020YbBaBO3}. Similarly, in the corresponding terbium-based system, the absence of magnetic Bragg or diffuse scattering down to $75\text{ mK}$ points to the suppression of three-dimensional long-range order (LRO), underpinned by a remarkably high frustration parameter of $f\gtrsim 95$ \cite{Kelly2025Tb}. These signatures, like the absence of LRO, large frustration index and coexisting ferromagnetic and antiferromagnetic exchanges, make them suitable candidates to be explored for MCE.

For Dy and Ho members of this family, we provide a comprehensive magnetic study through isothermal magnetization measurements down to $T \sim 1.8\text{ K}$ in applied magnetic fields up to $H = 70\text{ kOe}$. Both compounds crystallize in a hexagonal structure with space group $P6_3cm$, wherein nonmagnetic $\mathrm{Ba^{2+}/B^{3+}}$ structural units separate two-dimensional triangular planes of $\mathrm{RE^{3+}}$ ions, yielding a geometrically frustrated, quasi-two-dimensional triangular lattice. Temperature-dependent DC magnetic susceptibility confirms the absence of long-range ordering down to the lowest measured temperatures. Isothermal magnetization curves were analyzed to quantify the isothermal magnetic entropy change ($\Delta S_m(T, H)$). The given results provide fundamental insights into the magnetocaloric response of these compounds and demonstrate their potential applicability for adiabatic demagnetization refrigeration.

\section{Experimental methods}

\subsection{Synthesis}

Polycrystalline Ba$_3$RE(BO$_3$)$_3$ (RE= Dy, Ho) were prepared via a high-temperature solid-state reaction technique using rare-earth oxides (Dy$_2$O$_3$ and Ho$_2$O$_3$), BaCO$_3$, and H$_3$BO$_3$ as base materials \cite{Gao2018}. All the high-purity materials used here were purchased from Merck. Firstly, rare earth oxides were preheated at 800$^\circ$C for 12 hours (h), to remove any adsorbed moisture. Afterwards, respective rare earth oxides (Dy$_2$O$_3$ and Ho$_2$O$_3$), BaCO$_3$ and H$_3$BO$_3$ (5\% extra) were thoroughly mixed in stoichiometric amounts in an agate mortar and kept for heating at 500$^\circ$C for 12 h. Subsequently, the mixture was reground and pressed into pellets, then kept for heating at 950$^\circ$C for 24 h. With repeated grinding, the pellet were heated at the same temperature (950$^\circ$C) to obtain a single-phase compound.    

Morphological imaging of the prepared samples was performed using a scanning electron microscope (SEM) (Thermo Scientific Apreo 2S). Room-temperature powder X-ray diffraction (XRD) patterns were recorded using a powder X-ray diffractometer (Empyrean, PANalytical) with incident radiation of Cu K$\alpha$ = 1.540 Å at an operating voltage and current of 45 kV and 40 mA, respectively. The XRD data were performed over a $2\theta$ range of $10^\circ$-$70^\circ$. Rietveld refinement of the XRD pattern was performed using FullProf software \cite{FULLPROF} to get the crystallographic parameters and identify the constituent phases. Raman spectra of the prepared sample were obtained using a Raman spectrometer (Renishaw), with an incident source of 532 nm, grating 2400 Grooves/mm, and a charge-coupled device (CCD) as a detector. The X-ray photoelectron spectroscopy (XPS) data were collectd using the PHI Genesis XPS system. Al K$\alpha$ was used as an X-ray source with an energy of 1486.6 eV. Magnetic measurements were performed using a superconducting quantum interference device (SQUID) magnetometer (Quantum Design Magnetic Property Measurement System, MPMS-3). The measurements were conducted over a temperature range from 1.8--310~K under applied magnetic fields up to (70~kOe).

\begin{table}
\scriptsize\addtolength{\tabcolsep}{-1pt}
\caption{\label{table1}Structural parameters of Ba$_3$RE(BO$_3$)$_3$, with their respective units.}
\centering
\begin{ruledtabular}
\begin{tabular}{l c  r}

SG: & $P6_3cm$ \\
SG no.: & 185\\
RE= &Dy & Ho \\
{\it a=b}(\r{A}): & 9.4301(3)& 9.4278(3)\\
{\it c}(\r{A}): & 17.6483(2)& 17.6153(2)\\
{\it V}(\r{A}$^3$): & 1,359.16(2)& 1,355.96(2)\\
\hline
$l_{inter}$ & 8.825(2)& 8.808(2)\\
$l_{intra}$ & 5.445(2)& 5.443(2)\\
\end{tabular}
\end{ruledtabular}
\end{table}

\section{Results and Discussion}

\subsection{X Ray Diffraction}

The room temperature Rietveld refined XRD pattern of the synthesized Ba$_3$Dy(BO$_3$)$_3$ and Ba$_3$Ho(BO$_3$)$_3$ samples were depicted in Figure~\ref{Str}(a) and (b), respectively. The experimentally obtained Bragg diffraction peaks($I_{\text{obs}}$, black open circles) show excellent agreement with the calculated profiles ($I_{\text{cal}}$, red lines). The refinement confirms the hexagonal crystal structure for both Dy and Ho compounds, similar to a previously reported literature \cite{Gao2018, khamaganova1999}. The minimal intensity in the residual difference curves ($I_{\text{obs}} - I_{\text{cal}}$, blue lines) confirms the quality of the fit and validates the single-phase nature of both samples, with all observed diffraction peaks matching the expected Bragg reflections (indicated by magenta vertical ticks). The calculated refinement parameters are $R_{\mathrm{p}}$ = 4.23, $R_{\mathrm{wp}}$= 5.84 and $\chi^2$= 3.31 for Ba$_3$Dy(BO$_3$)$_3$ compound whereas $R_{\mathrm{p}}$= 6.07, $R_{\mathrm{wp}}$= 7.97 and $\chi^2$= 3.68 for Ba$_3$Ho(BO$_3$)$_3$ compound. The lattice parameters of both samples are given in Table I. These values are similar to the reported values of lattice parameters \cite{Gao2018,khamaganova1999}. In addition to the main phase, a tiny peak at $2\theta = 29.1^\circ$ (marked with a star in the XRD pattern) is observed, attributed to the residual RE$_2$O$_3$. Since the secondary phase constitutes only a very small fraction ($\leq 3\%$), its contribution is expected to be negligible and does not affect the conclusions of this study.

\begin{figure}
\includegraphics[width=0.9\linewidth]{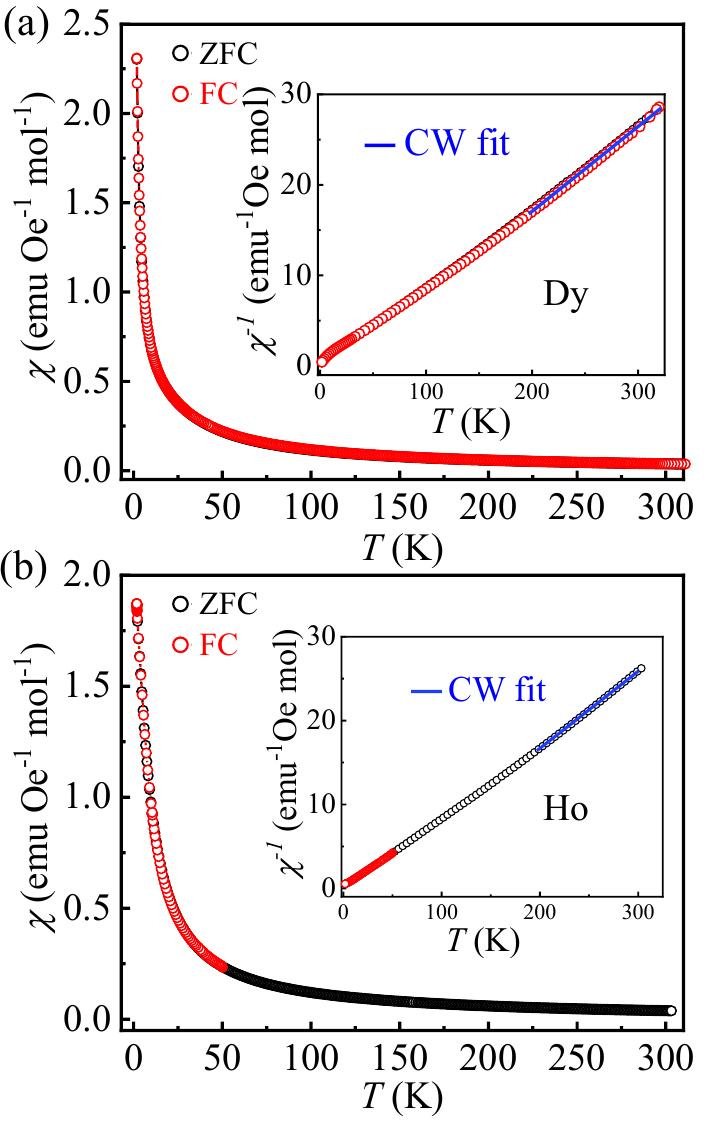}
\caption{(a) Temperature-dependent DC magnetic susceptibility ($\chi$) under zero-field-cooled (ZFC, black circles) and field-cooled (FC, red circles) conditions for \textbf{(a)} Ba$_3$Dy(BO$_3$)$_3$  and \textbf{(b)} Ba$_3$Ho(BO$_3$)$_3$. The insets show the inverse magnetic susceptibility ($\chi^{-1}$) as a function of temperature ($T$), where the solid blue line represents the Curie-Weiss (CW) linear fit at high temperatures.}
\label{MT}
\end{figure}
\begin{figure*}
\includegraphics[width=0.9\linewidth]{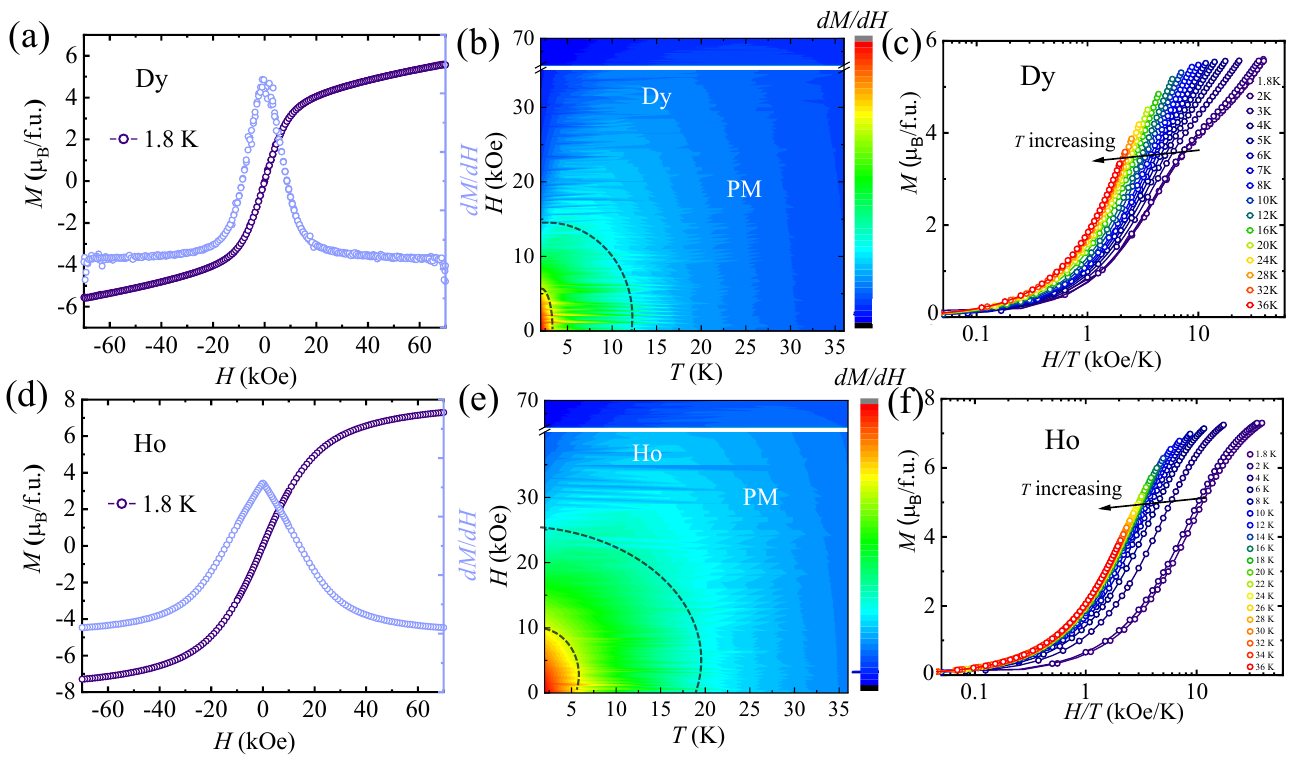}
\caption{\textbf{(a), (d)} Field-dependent magnetization ($M(H)$) curves (dark blue open circles, left axis) measured at $T = 1.8\text{ K}$ and the corresponding differential susceptibility $dM/dH$ (light blue open circles, right axis) for \textbf{(a)} Ba$_3$Dy(BO$_3$)$_3$ and \textbf{(d)} Ba$_3$Ho(BO$_3$)$_3$ up to $\pm 70\text{ kOe}$. \textbf{(b), (e)} Colour-contour phase diagrams of the magnetic field versus temperature ($H\text{--}T$) mapped using $dM/dH$ values for \textbf{(b)} Ba$_3$Dy(BO$_3$)$_3$ and \textbf{(e)} Ba$_3$Ho(BO$_3$)$_3$ compounds, depicting the paramagnetic (PM) region and field-induced crossover behavior (marked by dashed lines), \textbf{(c)} \textit{M} vs. \textit{H/T } plots of Ba$_3$Dy(BO$_3$)$_3$, and \textbf{(f)} Ba$_3$Ho(BO$_3$)$_3$}
\label{MH}
\end{figure*}
The refined structural models are illustrated in Fig.~\ref{Str}(c) and (d) using the VESTA package \cite{VESTA}.
The crystal unit cell can be viewed as an alternating stack of BaO and REO layers along the c-axis, having interlayer distances ($l_{inter}$) between RE ions of 8.825~\r{A} and 8.808~\r{A} for Dy and Ho compounds, respectively. Such a significant distance is ideal for having 2D-like magnetic exchange, a feature that is currently in high demand. Within each layer in the ab-plane, the rare-earth ions form a triangular lattice, with intralayer interatomic distances ($l_{intra}$) of 5.4452~\r{A} and 5.4434~\r{A} for Dy and Ho, respectively. These seemingly identical RE ions are different in Wyckoff positions, occupying 2a and 2b positions. The large difference between $l_{inter}$ and $l_{intra}$ suggests enhanced 2D magnetic excitations. The structural features and characteristic interatomic distances/layer spacings ($l_{inter}$ and $l_{intra}$) along distinct projections are delineated in Figure~\ref{Str}(c) and \ref{Str}(d), respectively. SEM images of both samples, shown in Figure~$S1$ of the Supplementary Information, reveal a microstructure consisting of relatively large particles with a crystal-like morphology. XPS data of both compounds are presented in Figure~$S2$ of the Supplementary Information, which reveal the presence of the rare earth elements (Dy and Ho) along with oxygen (O), carbon (C), barium (Ba), and boron (B) elements. 

\subsection{Temperature-dependent magnetic susceptibility}

The temperature-dependent magnetic susceptibility ($\chi(T)$) was recorded under both zero-field-cooled (ZFC) and field-cooled (FC) protocol modes across the temperature span of $1.8\text{ K}$ to $310\text{ K}$, for $H=$ 1~kOe. Figure~\ref{MT}(a) and \ref{MT}(b) present the $\chi(T)$ data for the Ba$_3$Dy(BO$_3$)$_3$ and Ba$_3$Ho(BO$_3$)$_3$, respectively. For both compounds, $\chi(T)$ increases monotonically as temperature decreases, exhibiting characteristic paramagnetic behavior at higher temperatures followed by a sharp, steep rise in $\chi(T)$ below $\sim 50\text{ K}$. No peak-shaped signature in low-temperature $\chi(T)$ was observed, which further shows the absence of long-range magnetic ordering. The ZFC and FC $\chi(T)$ curves overlap seamlessly across the measured temperature span, showing no discernible bifurcation or magnetic hysteresis, which further confirms the absence of long-range ferromagnetic ordering or spin-freezing effects down to $1.8\text{ K}$. These features are consistent with magnetically frustrated triangular rare-earth lattices \cite{Kelly2025Tb, CHEN2025} and consistent with the previous report on magnetic properties of Ba$_3$RE(BO$_3$)$_3$ compounds \cite{Gao2018}.
\begin{figure*}
\includegraphics[width=0.8\linewidth]{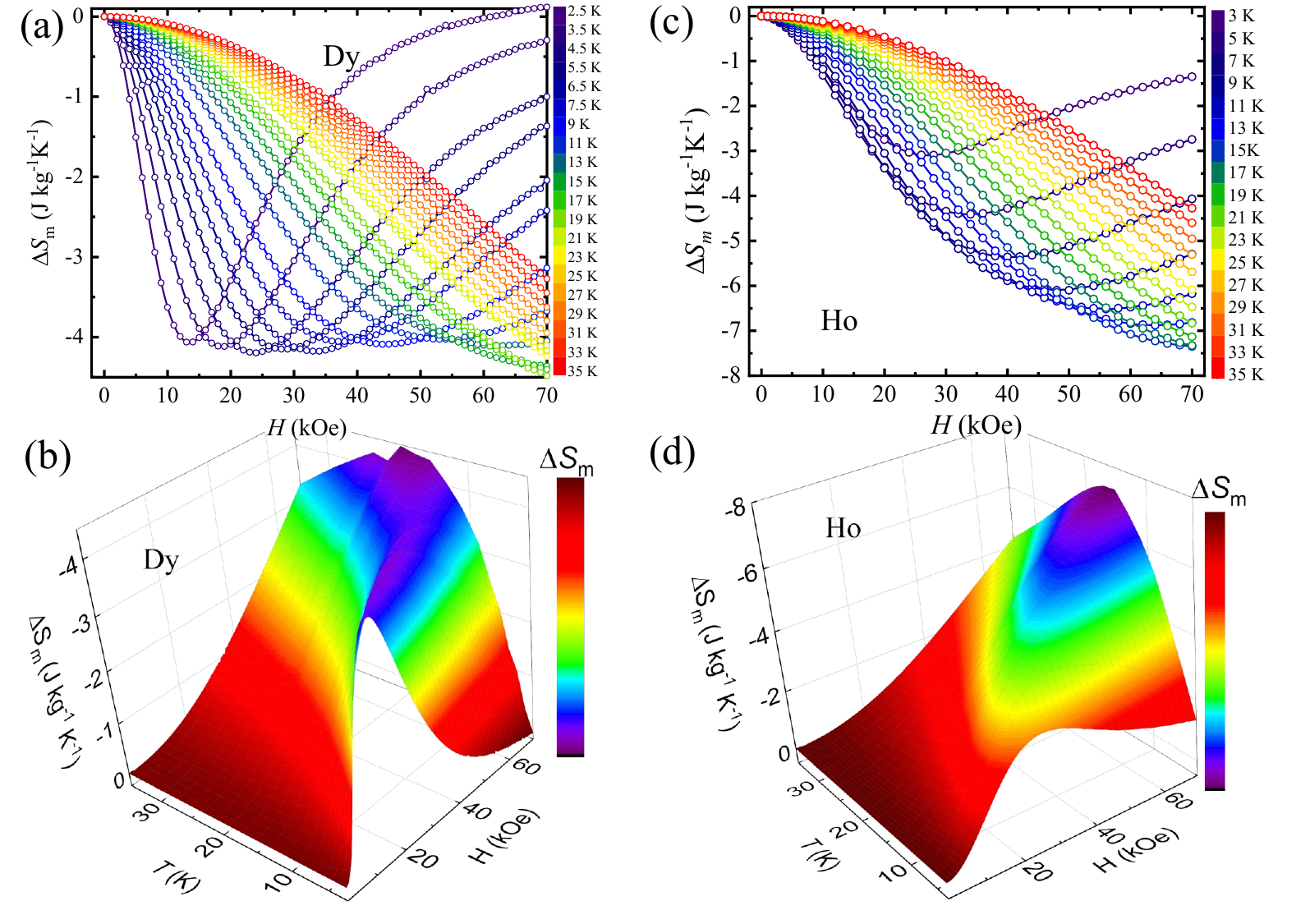}
\caption{ \textbf{(a), (c)} Magnetic field dependence of the isothermal magnetic entropy change ($\Delta S_m$) at various selected temperatures ranging from $2.5\text{ K}$ to $35\text{ K}$ up to a maximum applied magnetic field of $70\text{ kOe}$ for the \textbf{(a)} Dy- and \textbf{(c)} Ho-based compounds. 
\textbf{(b), (d)} Three-dimensional (3D) surface plots representing the magnetic entropy change ($\Delta S_{\text{m}}$) as a function of both magnetic field ($H$) and temperature ($T$) for \textbf{(b)} Dy and \textbf{(d)} Ho, highlighting the maximum entropy change at low temperatures and high magnetic fields.}
\label{MCE}
\end{figure*}
Further, to analyze the localized magnetic moments and exchange interactions, the high-temperature inverse magnetic susceptibility $\chi^{-1}(T)$ was fitted using the Curie-Weiss (CW) law:
\begin{equation}
\chi(T) = \frac{C}{T - \theta_{\text{CW}}}
\end{equation}
where $C$ is the Curie-Weiss constant and $\theta_{\text{CW}}$ is the Curie-Weiss temperature.
As shown in the insets of Figure~\ref{MT}(a) and \ref{MT}(b), 
the inverse susceptibility ($\chi^{-1}(T)$) demonstrates a linear dependence on temperature in the high-temperature regime ($190\,\mathrm{K} \lesssim T \lesssim 300\,\mathrm{K}$). The solid blue lines represent the linear Curie-Weiss fits to the experimental $\chi^{-1}(T)$ data, shown in the respective figures. The obtained Curie-Weiss temperatures ($\theta_{\text{CW}}$)  are $18\text{ K}$ and $20 \text{ K}$ for Dy and Ho compounds, respectively. The effective magnetic moments ($\mu_{\text{eff}}$) extracted from the slope ($C$) of the fits are $\mu_{\mathrm{eff}} = 9.31\,\mu_{\mathrm{B}}$ for Dy and $\mu_{\mathrm{eff}} = 9.30\,\mu_{\mathrm{B}}$ for Ho, which is close to the theoretical free-ion values for $\text{Dy}^{3+}$ ($J = 15/2, \, g_J = 4/3, \, \mu^{\text{Dy}^{+3}}_{\text{eff}} (\text{theo})\approx 10.65\,\mu_{\text{B}}$) \cite{Hutchison2016} and 
$\text{Ho}^{3+}$ ($J = 8, \, g_J = 5/4, \, \mu ^{\text{Ho}^{+3}}_{\text{eff}}(\text{theo}) \approx 10.60\,\mu_{\text{B}}$) \cite{Blundell2001}. 
The extrapolated $\theta_{\text{CW}}$ provide insight into the nature and strength of the magnetic exchange coupling among magnetic moments. Relatively large values of $\theta_{\text{CW}}$ and the absence of long-range magnetic ordering up to 2~K suggest significant magnetic frustration, consistent with other sister compounds \cite{Kelly2025Tb}. The large frustration index and absence of LRO are commonly observed in other compounds of this triangular-net family \cite{Gao2018, Zeng2020YbBaBO3, ennis2024_Ba3ErBO3, Kelly2025Tb}.

\subsection{Isothermal magnetization}

To investigate the field-induced magnetic behavior at low temperatures, field-dependent magnetization ($M(H)$) were measured at $1.8\text{ K}$ in the field range of $\pm70~$kOe. 
Figure~\ref{MH}(a) and \ref{MH}(d) depict the field-dependent magnetization $M(H)$ along with the differential susceptibility $dM/dH$ at $1.8\text{ K}$ for the Ba$_3$Dy(BO$_3$)$_3$ and Ba$_3$Ho(BO$_3$)$_3$ compound, respectively. Both systems exhibit standard \textit{S}-shaped, non-hysteretic magnetization curves, indicative of field-polarised states without remanent magnetization or coercive fields, confirming the absence of long-range magnetic ordering. For the Dy compound, $M(H)$ increases steadily with the applied field up to 20~kOe, followed by a nearly linear increase with field, reaching approximately $5.6~\mu_{\mathrm{B}}/\mathrm{f.u.}$ at 70~kOe. In contrast, the $\text{Ho}$ compound exhibits a relatively smoother slope shift around 30~kOe. It attains a higher magnetic response, achieving $\sim 7.3\,\mu_{\text{B}}/\text{f.u.}$ under the maximum applied field of $70\text{ kOe}$. The experimental saturation magnetization values from high-field $M(H)$ measurements, are noticeably reduced compared to their theoretical free-ion limit of ($g_JJ=$) $10.0\ \mu_\text{B}/\text{ion}$ owing to crystal field splitting of $f$-states and magnetocrystalline anisotropy constraints \cite{Guo2019KBaREBO3, Rooney2026MgLnB5O10}. 
Furthermore, the corresponding differential susceptibility curves $dM/dH$ (on the right y-axis) display symmetric, single-peaked behavior centered at zero field ($H = 0$), gradually decreasing toward higher fields as the spin moments approach alignment with increasing magnetic field. 

Furthermore, multiple $M(H)$ measurements were performed (0 $\rightarrow$ 70 kOe) for Ba$_3$Dy(BO$_3$)$_3$ and Ba$_3$Ho(BO$_3$)$_3$ at temperatures ranging from 2 to 36 K, and are presented in Figure $S4(a)$ and $S4(b)$ in the Supplementary Information, respectively. For both compounds, the $M(H)$ curves show nonlinear behaviour in the applied magnetic field at temperatures up to 20~K, suggesting deviation from pure paramagnetism in the low-temperature regime. This observation further supports the occurrence of magnetic correlation established without a long-range ordered phase \cite{basu2017, Rooney2026MgLnB5O10}.
To map out the broad phase boundaries across temperature and magnetic field space, contour plots of $dM/dH$ as a function of temperature ($T$) and magnetic field ($H$) are presented in Figure~\ref{MH}(b) and \ref{MH}(e) for $\text{Dy}$ and $\text{Ho}$ compounds, respectively. The color intensity mapping corresponds to the magnitude of $dM/dH$, where the highest differential susceptibility (red region) is concentrated at low fields ($H < 10\text{ kOe}$) and low temperatures ($T < 5\text{ K}$). This is further verified through $M$ vs. $H/T$ plots, shown in Figure~\ref{MH}(c) and \ref{MH}(f) for $\text{Dy}$ and $\text{Ho}$ compounds, respectively, where distinct isothermal curves deviate from the universal curve, suggesting a low-temperature correlated magnetic state rather than a pure paramagnetic one \cite{Jogendra2025TbHoO4}. The curves at relatively higher temperatures ($T>20~$K) fall on each other; above this temperature, magnetic interactions are paramagnetic-like.
Overall, the evolution of $dM/dH$ across $H$--$T$ space reveals strong spin polarization dynamics at low temperatures, with no abrupt field-induced metamagnetic transitions detected up to $70\text{ kOe}$. Additionally, we have shown the Arrott plots ($M^2$ vs. $M/H$) for both compounds in the Supplementary Information.

\subsection{Magnetocaloric effect}

Rare-earth-based materials, lacking long-range magnetic ordering, are promising candidates for magnetic refrigeration in the cryogenic temperature regime \cite{Guillou2014, Reynolds2019}, further motivating evaluation of the magnetocaloric effect and thermodynamic performance of these compounds. The isothermal magnetic entropy change ($\Delta S_{\text{m}}$) was calculated from isothermal $M(H)$ curves measured at various temperatures using the integrated Maxwell relation \cite{phan2007,MishraCOM2024}:

\begin{equation}
\Delta S_{\text{m}}(T, \Delta H) = \int_{0}^{H} \left( \frac{\partial M}{\partial T} \right)_{\!H} dH
\end{equation}

Figure~\ref{MCE}(a) and \ref{MCE}(c) illustrate the magnetic field dependence of $\Delta S_{\text{m}}$ at particular temperatures ranging from $2.5\text{ K}$ to $35\text{ K}$ up to a maximum applied field of $70\text{ kOe}$ for the $\text{Dy}$ and $\text{Ho}$ compounds, respectively. For both compounds, $-\Delta S_{\text{m}}$ increases gradually with increasing magnetic field over the entire temperature interval, reflecting field-induced alignment of localized moments. 
For both Dy and Ho compounds, $-\Delta S_{\text{m}}$ varies nonlinearly with applied field. 
At low temperatures ($T<10~$K), the entropy variation depicts a change in slope at the critical fields after achieving its maximum value, which is consistent with isothermal $M(H)$ data. The observed critical field shifted to higher values with an increase in temperature. For the $\text{Dy}$ compound, a maximum entropy change of approximately $-\Delta S_{\text{m}} \approx 4.2\text{ J kg}^{-1}\text{ K}^{-1}$ is observed at 24~kOe, whereas the $\text{Ho}$ counterpart exhibits a significantly larger maximum value of $-\Delta S_{\text{m}} \approx 7.3\text{ J kg}^{-1}\text{ K}^{-1}$ under $\Delta H = 70\text{ kOe}$. 
The magnetic entropy in these rare-earth systems is governed by local crystal electric field splitting of the Hund's rule ground manifold; because the excitation gap ($\Delta_\text{CEF}$) vastly exceeds laboratory Zeeman energies ($\mu_\text{B}H$), higher-lying states remain inaccessible, capping the isothermal entropy change at the doublet limit of $\Delta S_\text{m} \approx R \ln 2 \approx 5.763 \text{ J mol}^{-1}\text{K}^{-1}$ (or $\approx 7.67 \text{ J kg}^{-1}\text{K}^{-1}$ depending on molar mass normalization), with experimental reductions below this theoretical limit stemming from polycrystalline averaging and magnetic frustration \cite{Rooney2026MgLnB5O10}. Within this framework, the distinct behavior of Dy- and Ho-based compounds arises from their electronic configurations and time-reversal symmetry: as a $4f^9$ Kramers ion, Dy$^{3+}$ possesses a protected ground doublet yielding a linear Zeeman response, whereas the $4f^{10}$ non-Kramers Ho$^{3+}$ ion frequently exhibits a singlet ground state governed by quadratic Van Vleck mixing \cite{Zeng2024DyV6Sn6}. Consequently, while the Dy compound exhibits a smaller absolute entropy change magnitude, its enhanced low-field sensitivity shifts $\Delta S_\text{max}$ to significantly lower applied fields, establishing it as an optimal candidate for low-field magnetocaloric refrigeration. For both compounds, plots for $\Delta S_m$ vs. $T$ are shown in the supplementary information (see Figure $S6$).

To further elucidate the evolution across temperature and field, three-dimensional (3D) surface projections of $\Delta S_{\text{m}}(T, H)$ are depicted in Figure~\ref{MCE}(b) and \ref{MCE}(d) for $\text{Dy}$ and $\text{Ho}$ compound, respectively. The 3D landscape clearly highlights a pronounced ridge centered around the low-temperature regime ($T < 10\text{ K}$), where spin entropy suppression is maximized under an applied magnetic field. The smooth variation of $\Delta S_{\text{m}}$ with increasing temperature is consistent with second-order, reversible magneto-caloric behavior, rendering these materials promising candidates for low-temperature cryogenic refrigeration applications. Based on the comparative investigation, we found that the Dy compound is a better magneto-cooling material in low applied fields ($H<10$~kOe).

\begin{figure}[t]
\includegraphics[width=0.9\linewidth]{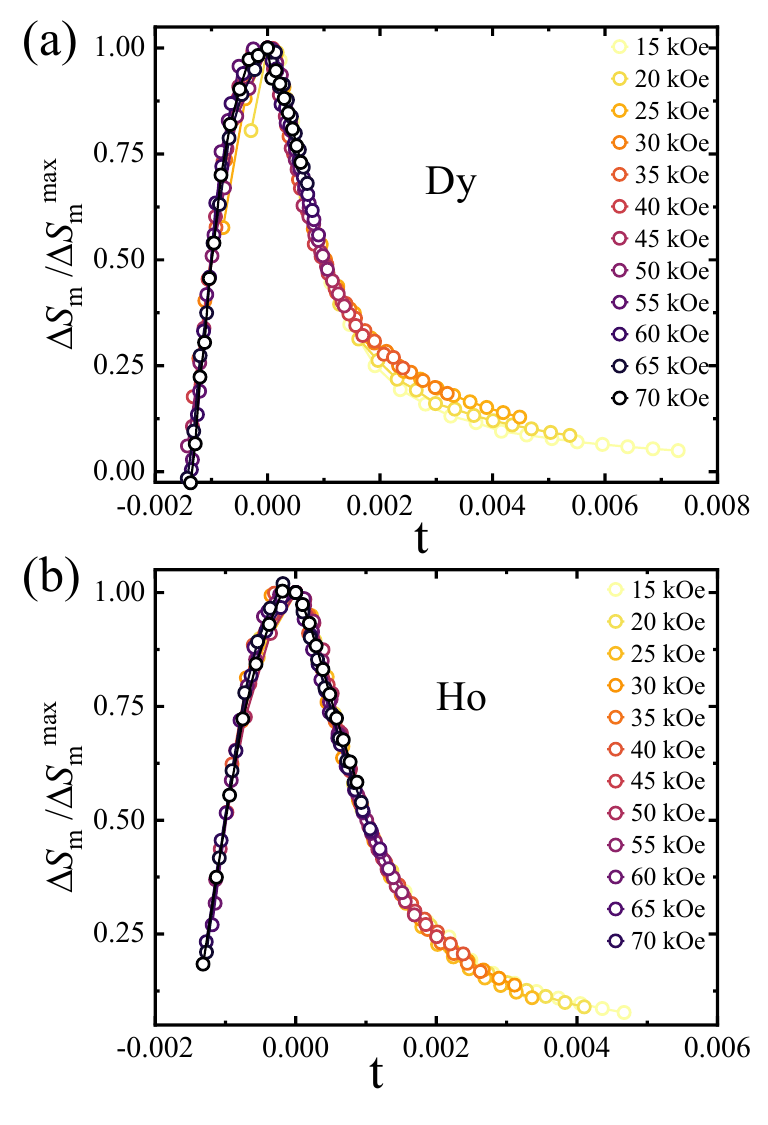}
\caption{Normalized magnetic entropy change, $\Delta S_m/\Delta S_m^{\mathrm{max}}$, as a function of reduced temperature $t$ for (a) Ba$_3$Dy(BO$_3$)$_3$ and (b) Ba$_3$Ho(BO$_3$)$_3$ at different applied magnetic fields.}
\label{MPD}
\end{figure}

Furthermore, the scaling behaviour can be formulated by normalising all the $\Delta S_m$ curves with their corresponding maximum values ( i.e., $\Delta S_{\mathrm{m}}/\Delta S_{\mathrm{m}}^{\max}$) for each magnetic field change and plotted as a function of reduced temperature ($t$), given in Figure~\ref{MPD}. The reduced temperature ($t$) is calculated using the given formula below\cite{Yang2021PRB, FRANCO2010}:

\begin{equation}
    t_- =  (T_{peak}-T)/(T_{r1}-T_{peak}),    T\leq T_{peak} 
\end{equation}

\begin{equation}
    t_+ =  (T-T_{peak})/(T_{r2}-T_{peak}),    T\geq T_{peak}  
\end{equation}
\begin{figure*}
\includegraphics[width=.8\linewidth]{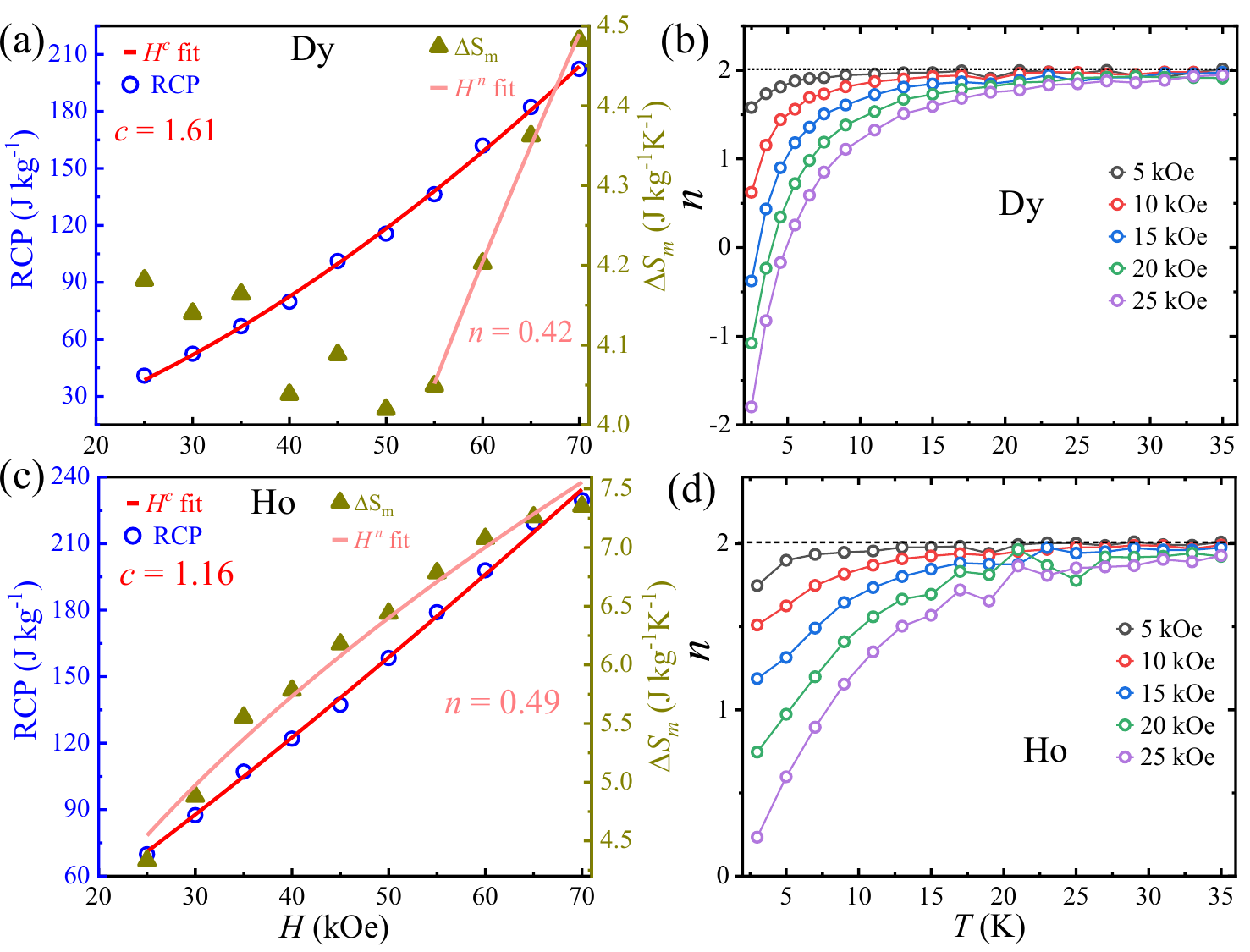}
\caption{The variation in Relative cooling power (RCP) and  $\Delta S_m$ as a function of applied field
($H$), solid red line shows an $H^c$ fit, and the light red line shows an $H^n$ fit for (a) Dy and (c) Ho compounds. Variation in $n$ as a function of temperature at various applied fields, for (b) Dy and (d) Ho compounds.}
\label{RCP}
\end{figure*}
The normalised magnetic entropy change, $\Delta S_{\mathrm{m}}/\Delta S_{\mathrm{m}}^{\max}$, with respect to the reduced temperature ($t$) is shown in Figure \ref{MPD}(a) and \ref{MPD}(b) for the Dy- and Ho-based compounds, respectively. The curves obtained under different magnetic-field changes ($\Delta H = 10$--70~kOe, in 5~kOe increments) depict a nearly universal scaling behaviour, with the curves closely overlapping over a broad temperature span. Such field-independent scaling of the normalised entropy-change curves is a characteristic feature of a second-order magnetic behaviour \cite{bonilla2010_MPT}. 

The key parameter for evaluating the magnetocaloric efficiency of any material is the relative cooling power (RCP), which estimates the heat transferred between the cold and hot reservoirs during a refrigeration cycle. It is defined as the integral of the absolute magnetic entropy change over the operating temperature span, which can represented as

\[
\mathrm{RCP}=\int_{T_{\mathrm{cold}}}^{T_{\mathrm{hot}}}
|\Delta S_m(T,H)|\,dT,
\tag{5}
\]

where \(T_{\mathrm{cold}}\) and \(T_{\mathrm{hot}}\) depict the temperatures of the cold and hot systems, respectively, and \(\Delta S_m(T,H)\) expresses the magnetic entropy variation as a function of temperature and applied magnetic field. The RCP can be approximated as \cite{wood1985, Murtaza2020}

\[
\mathrm{RCP}\approx|\Delta S_m^{\max}|
\times\delta T_{\mathrm{FWHM}}
\tag{6}
\]

where \(|\Delta S_M^{\max}|\) is the maximum absolute magnetic entropy change and \(\delta T_{\mathrm{FWHM}}\) is the corresponding temperature width at half maximum. The maximum RCP values found to be ~204, and ~230 $\mathrm{J\,kg^{-1}}$ at 70~$\mathrm{kOe}$ for the Ba$_3$Dy(BO$_3$)$_3$ and Ba$_3$Ho(BO$_3$)$_3$, respectively. The obtained RCP values are comparable to those of other magnetocaloric materials, as tabulated in Table \ref{tab:Dy_compounds_MCE}.

Magnetic refrigeration relies on a cyclic process of continuous magnetization and demagnetization. Materials that exhibit first-order phase transitions are unsuitable for cyclic processes due to significant magnetic or thermal hysteresis losses \cite{franco2012RCP}. Instead, materials that show second-order phase transitions are more desirable and better suited for magnetic refrigeration. To understand the nature of a material's phase transition, we analyse the field dependence of its RCP \cite{islam2020}, as shown in Figure~\ref{RCP}. A power-law fit of the form RCP$\propto H^c$
performed on the RCP($H$) curve for both Dy- and Ho-based compounds, as given in the corresponding Figure \ref{RCP}(a) and \ref{RCP}(c), respectively. The values of c are estimated to be $\sim 1.61$ and $\sim 1.16$ for Dy- and Ho-based compounds, respectively. Similarly, the exponent for $\Delta S_m$ vs. $H$, defined as $n$ is found to be $\sim 0.42$ (Dy) and $\sim 0.49$ (Ho). The exponent $n$ is a field- and temperature-dependent parameter and can also be calculated locally as
\begin{equation}
    n(T, H) = \frac{d \ln |\Delta S_{\text{m}}|}{d \ln H}
    \tag{7}
\end{equation}
.

The obtained values for both compounds are shown in Figure \ref{RCP}(b) and (d). In the high-temperature limit, $n$ is close to 2, as usually observed in the paramagnetic regime; as the temperature decreases, $n$ decreases gradually without a minimum. The observed feature implies a possible transition upon further lowering the temperature in the millikelvin range. The overall behaviour of $n$ vs $T$ curves (in the explored temperature regime $T>2~$K) is found to be similar to other compounds showing a second-order phase transition \cite{islam2020, moreno2021_nT}, consistent with Arrott plot analysis, as shown in the Supplementary Information. 





\begin{table*}[btp]
\centering
\caption{Comparison of magnetic transition temperature ($T_N$/$T_M$), magnetocaloric entropy change ($\Delta S_M^{\max}$), and relative cooling power (RCP) for representative compounds.}
\label{tab:Dy_compounds_MCE}

\renewcommand{\arraystretch}{1.2}

\begin{tabular}{lcccc}
\hline\hline

Compound &
$T_N$/$T_M$ (K) &
$\Delta S_M^{\max}$ (J\,kg$^{-1}$K$^{-1}$) &
RCP (J\,kg$^{-1}$) &
References \\

\hline

Ba$_3$Dy(BO$_3$)$_3$ & -- & 4.2 &
204 & This work \\

DyBO$_3$ &
$T_N < 2.8$ & 13.9 &
--- & \cite{MUKHERJEE2018} \\

LiDyP$_4$O$_1$$_2$ &-- &
$\sim 11.7$ & 236 & \cite{petrov2020}\\

Gd$_{0.9}$Dy$_{0.1}$PO$_4$ &
*PT ($T_{\mathrm{M}}$) = 2.6~K &
$\sim 43.7$ &
$\sim 262.4 $ & \cite{fang2026}\\

DyVO$_4$ & $T_N=3.5$ &
26 &
RC = 526 & \cite{MIDYA2014}\\

Gd$_0.8$Dy$_0.2$PO$_4$ &
*PT ($T_{\mathrm{M}}$) = 2.4~K &
41 &
--- & \cite{fang2026}\\

DyCrTiO$_5$ &
$T_N \approx 153$, *SR $\approx$ 49 &
$\sim 10.9$ &
--- & \cite{Bamana2023}\\

Dy$_3$NbO$_7$ &
--- &
$ 19.39$ &
$ 444.39 $ & \cite{chen2025_table}\\

DyNaGeO$_4$ &
--- &
$ 14.80$ &
$ 236.82 $ & \cite{zhang2025_table}\\

\hline
Ba$_3$Ho(BO$_3$)$_3$ & -- & 7.3 &
230 & This work \\

HoPO$_4$ &
$T_N < 2.4$ &
$34.5 $ & $414.8 $ & \cite{elouafi2024} \\

Ho$_2$Ti$_2$O$_7$ & $T_N < 2$ & 
13.4 &
RC = 86.0 & \cite{Xu2024}\\

HoVO$_4$ & -- &
7.94 &
201.3 & \cite{dey2017}\\

HoBO$_3$ & &
0.9 &
--- & \cite{MUKHERJEE2018}\\

HoTiO$_3$ & 53 &
5.96 &
111.7 & \cite{Su2011} \\

Ho$_3$NbO$_7$ &
--- &
$ 18.52$ &
$ 495.9 $ & \cite{chen2025_table}\\

HoNaGeO$_4$ &
--- &
$ 15.370$ &
$ 239.47 $ & \cite{zhang2025_table}\\

\hline\hline
\end{tabular}

*SR= spin reorientation, PT= Phase transition
\end{table*}

\section{Conclusion}

In this work, polycrystalline Ba$_3$RE(BO$_3$)$_3$, where RE = Dy and Ho, were prepared by the solid-state reaction method. We systematically investigated the crystal structure, elemental identification, magnetic properties, and magnetocaloric effect. The $\chi(T)$ curves confirm no discernible bifurcation or magnetic hysteresis across the temperature span for both compounds, confirming the absence of long-range ferromagnetic ordering or spin-freezing effects down to $2\text{ K}$, while a sharp rise at low temperature suggests short-range correlation ($T<20$~K). For both compounds (i.e., Dy and Ho), the $M(H)$ isotherms show nonlinear behaviour up to 16 K, suggesting field-induced polarisation in the low-temperature regime. Interestingly, maximum entropy change of approximately $-\Delta S_{\text{m}} \approx 4.2\text{ J kg}^{-1}\text{ K}^{-1}$ is observed for the $\text{Dy}$ compound at 24~kOe, whereas the $\text{Ho}$ counterpart exhibits a significantly larger maximum value of $-\Delta S_{\text{m}} \approx 7.3\text{ J kg}^{-1}\text{ K}^{-1}$ under $\Delta H = 70\text{ kOe}$. In both compounds, the normalised magnetic entropy change, $\Delta S_{\mathrm{M}}/\Delta S_{\mathrm{M}}^{\max}$, as a function of the reduced temperature ($t$) shows universal scaling behaviour, consistent with a second-order magnetic phase transition. The maximum relative cooling power (RCP) values were found to be ~204 (Dy) and ~230 (Ho) $\mathrm{J\,kg^{-1}}$ at 70~$\mathrm{kOe}$. These results open new avenues for exploring rare-earth-based triangular-net lattices Ba$_3$RE(BO$_3$)$_3$ (RE = Dy, Ho) as potential materials for low-temperature magnetic refrigeration, particularly in applications for cryogenic refrigeration.

\section*{ACKNOWLEDGMENTS}
KNP and SBS sincerely acknowledge the Central Advanced Instrumental Facility at IISER Berhampur for XRD, XPS, and MPMS-3 facilities. KNP further acknowledges Harshvardhan from IISER Berhampur for his assistance in Rietveld refinement. 

\bibliography{REBBO_cleaned}

\end{document}


\begin{titlepage}
\centering
\renewcommand{\thefootnote}{\fnsymbol{footnote}}

\vspace*{4cm}

{\Huge\bfseries Supplementary Information\par}
\vspace{1.5cm}

{\LARGE\bfseries 
Competing interaction driven  broad magnetic entropy distribution in triangular net compounds \text{Ba$_3$RE(BO$_3$)$_3$},  RE= Dy and Ho
\par}
\vspace{1.5cm}
{\normalsize
\textbf{Krishna Nand Prajapati},$^{1}$\footnote{E-mail: knpiitb@gmail.com}
\textbf{Prabuddha Kant Mishra},$^{2}$
\textbf{Shashi B Singh},$^{1}$\footnote{E-mail: sbsingh@iiserbpr.ac.in}
\par}
\vspace{1cm}

{\small

$^{1}$ Department  of Physical Sciences, Indian Institute of Science Education and Research, Berhampur, Odisha 760003, India\\[0.3cm]

$^{2}$Central Advanced Instrumentation Facility, Indian Institute of Science Education and Research, Berhampur, Odisha 760003, India\\[0.3cm]

\par}

\vfill

\end{titlepage}

\setcounter{page}{2}
\pagestyle{fancy}
\fancyhf{}
\rhead{\thepage}


\begin{figure*}
    \centering
    \includegraphics[width=0.75\linewidth]{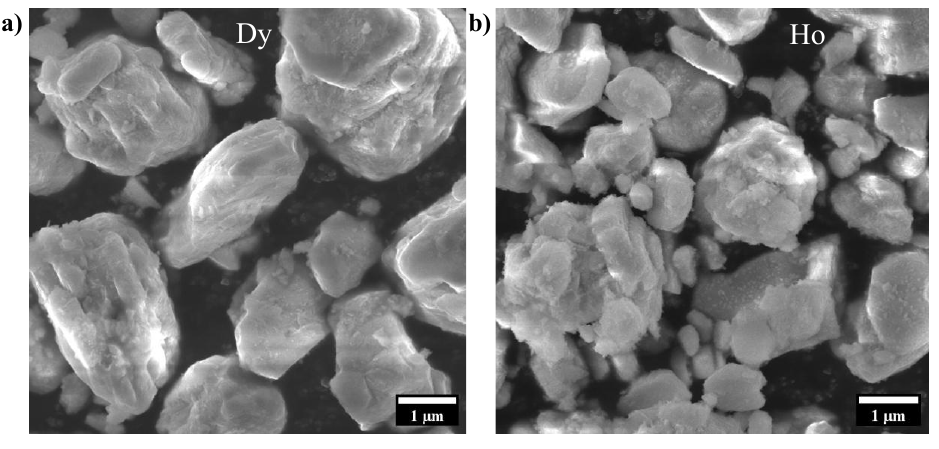}
    \caption{SEM image of polycrystalline specimen of (a) Ba$_3$Dy(BO$_3$)$_3$, (b) Ba$_3$Ho(BO$_3$)$_3$  .}
    \label{cry}
\end{figure*}
\begin{figure*}
    \centering
    \includegraphics[width=1\linewidth]{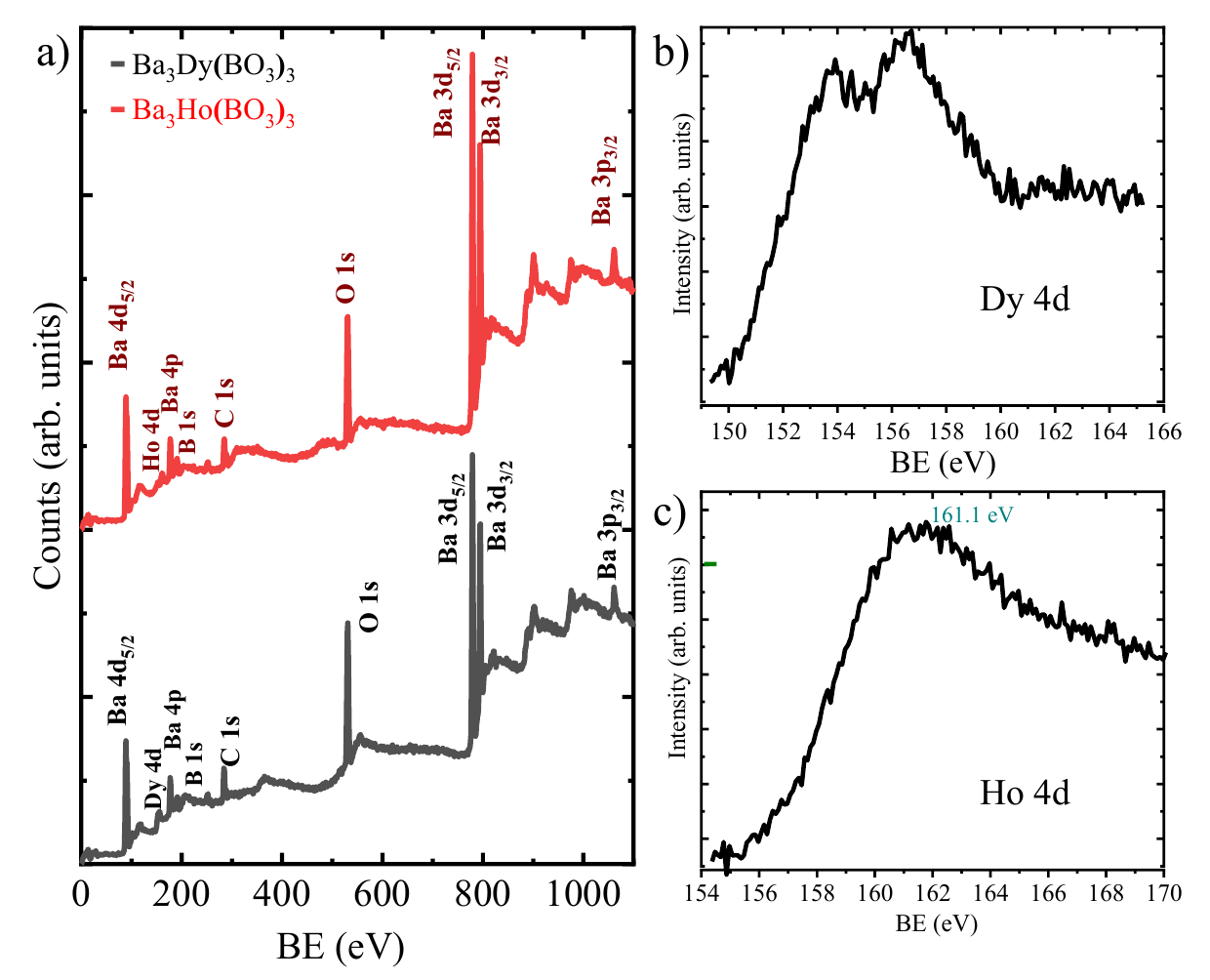}
    \caption{XPS spectra: (a) survey spectra of both Ba$_3$Dy(BO$_3$)$_3$ and Ba$_3$Ho(BO$_3$)$_3$  compounds; high-resolution spectra of (b) Dy~4d of Ba$_3$Dy(BO$_3$)$_3$ compound, (c) Ho~4d of Ba$_3$Ho(BO$_3$)$_3$ compound.}
    \label{dis}
\end{figure*}

\begin{figure*}
    \centering
    \includegraphics[width=0.6\linewidth]{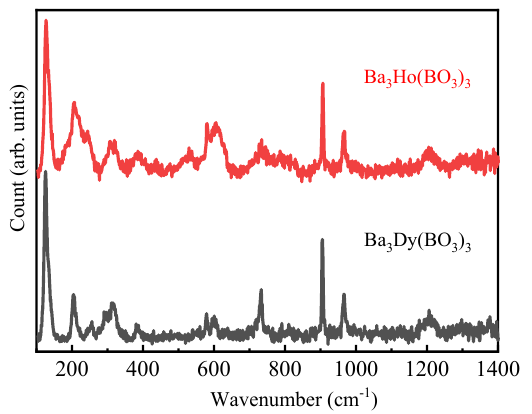}
    \caption{Raman spectra of Ba$_3$Dy(BO$_3$)$_3$, Ba$_3$Ho(BO$_3$)$_3$ .}
    \label{edx1}
\end{figure*}
\begin{figure}
\begin{center}
    \includegraphics[width=0.85\linewidth]{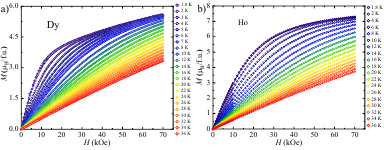}
    \caption{Isothermal magnetization ($M$) as a function of the applied magnetic field ($H$) measured at various temperatures ranging from 1.8 K to 36 K for: (a) Ba$_3$Dy(BO$_3$)$_3$, and (b) Ba$_3$Ho(BO$_3$)$_3$.}
    \label{fig3}
\end{center}
\end{figure}
\begin{figure}
\begin{center}
    \includegraphics[width=0.85\linewidth]{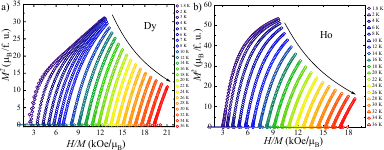}
    \caption{Arrott plots at various temperatures ranging from 1.8 K to 36 K for: (a) Ba$_3$Dy(BO$_3$)$_3$, and (b) Ba$_3$Ho(BO$_3$)$_3$.}
    \label{fig5}
\end{center}
\end{figure}

\begin{figure}
\begin{center}
    \includegraphics[width=0.85\linewidth]{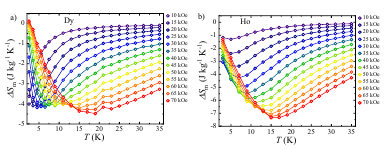}
    \caption{Temperature dependence of the magnetic entropy change ($\Delta S_\text{m}$) under various magnetic fields up to $70\text{ kOe}$ for: (a) Ba$_3$Dy(BO$_3$)$_3$, and (b) Ba$_3$Ho(BO$_3$)$_3$.}
    \label{fig6}
\end{center}
\end{figure}
\newpage

\section{XPS}
XPS spectra were recorded to investigate the surface composition and chemical state of both Ba$_3$Dy(BO$_3$)$_3$ and Ba$_3$Ho(BO$_3$)$_3$ compounds. The survey scans of both compounds are presented in Figure~$S2$(a), which corroborates the presence of rare earth elements (Dy and Ho), along with oxygen (O), carbon (C), barium (Ba), and boron (B). All the peaks are identified and highlighted in the survey spectra, as shown in Figure $S2(a)$. Figure~$S2$(b) shows the high-resolution Dy 4d core-level spectra for the Ba$_3$Dy(BO$_3$)$_3$ compound. The obtained spectra of Dy 4d are consistent with the previously reported Dy $4d$ binding energy, confirming the Dy$^{3+}$ oxidation state \cite{barreca2007_XPS, tholkappiyan2015_XPS}. Figure~$S2$(c) presents the core-level XPS spectra of Ho~4d of the Ba$_3$Ho(BO$_3$)$_3$ compound. The Ho~4d of the Ba$_3$Ho(BO$_3$)$_3$ compound exhibits a peak maximum at a binding energy of 161.1 eV, corresponding to the Ho~4d states\cite{rath2019_XPS, pan2014_XPS}. This characteristic binding energy confirms the presence of Ho$^{3+}$ ions in the compound.


\newpage
\bibliographystyle{unsrt}
\bibliography{REBBO}